# Propagation of spin-waves packets in individual nano-sized yttrium iron garnet magnonic conduits

Björn Heinz[1,2]*, Thomas Brächer[1], Michael Schneider[1], Qi Wang[1,3], Bert Lägel[4], Anna M. Friedel[1], David Breitbach[1], Steffen Steinert[1], Thomas Meyer[1,5], Martin Kewenig[1], Carsten Dubs[6], Philipp Pirro[1] and Andrii V. Chumak[1,3]

[1]Fachbereich Physik and Landesforschungszentrum OPTIMAS, Technische Universität Kaiserslautern, D-67663 Kaiserslautern, Germany.
[2]Graduate School Materials Science in Mainz, Staudingerweg 9, D-55128 Mainz, Germany.
[3]Faculty of Physics, University of Vienna, Boltzmanngasse 5, A-1090 Wien, Austria.
[4]Nano Structuring Center, Technische Universität Kaiserslautern, D-67663 Kaiserslautern, Germany.
[5]THATec Innovation GmbH, Augustaanlage 23, D-68165 Mannheim, Germany.
[6]INNOVENT e.V., Technologieentwicklung Jena, D-07745 Jena, Germany.

*Corresponding author
E-Mail: bheinz@rhrk.uni-kl.de

Modern-days CMOS-based computation technology is reaching its fundamental limitations. The emerging field of magnonics, which utilizes spin waves for data transport and processing, proposes a promising path to overcome these limitations. Different devices have been demonstrated recently on the macro- and microscale, but the feasibility of the magnonics approach essentially relies on the scalability of the structure feature size down to an extent of a few 10 nm, which are typical sizes for the established CMOS technology. Here, we present a study of propagating spin-wave packets in individual yttrium iron garnet (YIG) conduits with lateral dimensions down to 50 nm. Space and time resolved micro-focused Brillouin-Light-Scattering (BLS) spectroscopy is used to characterize the YIG nanostructures and measure the spin-wave decay length and group velocity directly. The revealed magnon transport at the scale comparable to the scale of CMOS proves the general feasibility of a magnon-based data processing.

## INTRODUCTION

The further scaling of CMOS-based computation technologies is increasingly costly and challenging due to several fundamental constrains, which arise from both, technological and physical limitations ([1]). To allow for further progress, the field of spintronics aims to complement CMOS by taking advantage of the spin degree of freedom to generate and control charge currents ([2]). However, a different approach to tackle these challenges and avoid electric currents can be found in the field of magnonics, which proposes a wave-based logic for more-than-Moore computing by utilizing spin waves to carry the information instead of electrons ([3–6]). The phase of a spin wave offers an additional degree of freedom enabling efficient computing concepts, which typically rely on the interference of coherent spin waves. In addition, spin waves possess readily accessible nonlinear phenomena which can be utilized to perform logic operations ([7–9]) and to realize novel computing devices. A variety of spin-wave based devices has already been realized on the macro- and microscale ([3, 5, 6, 9–18]), pushing partially into the nanoscale with structure sizes of a few 100 nm extent ([19–21]). However, a significant milestone on the path towards the development of magnonic circuits is the final advance to the sub-100 nm scale. In particular, yttrium iron garnet (YIG), the go-to material of magnonics which provides the lowest known spin-wave damping, is lacking this push to the nanoscale due to its complicated crystallographic structure ([22]) featuring a unit cell size of 1.2376 nm ([23]). Recently, we reported on the fabrication of sub-100 nm wide YIG nanostructures and the investigation of the uniform precession mode in them, developing an extended theoretical model describing the spin-wave dispersion in nanostructures (*24*), which takes an unpinning of the lateral mode profile for small aspect ratios into account. Nevertheless, neither propagating spin waves, which allow for the transport of information, nor the impact of the structuring process on their propagation have yet been investigated in structures of sub-100 nm lateral sizes.

In this letter, we report on the investigation of propagating spin waves in individual YIG magnonic conduits with lateral sizes down to 50 nm. Spin waves are excited continuously or pulsed using a micro-antenna and micro-focused Brillouin-Light-Scattering (BLS) spectroscopy is employed to directly measure the spin-wave decay length and group velocity. Considering a potential application in magnonic circuits the investigations are performed in the Backward-Volume (BV) geometry, which corresponds to the natural self-magnetized state of such a conduit to avoid the necessity of large bias magnetic fields. The results are compared to the theoretical predictions based on the extended model of the spin-wave dispersion in nanostructures (*24*). The measured spin-wave decay length is found to be in agreement with this theory, indicating only a moderate influence of the nanostructuring process on the YIG parameters. Moreover, the decay length in the smallest investigated structure is twice larger than the expected theoretical value of the excited dipolar spin waves, assuming a possible contribution of short-wavelength exchange waves to the detected BLS signal.

### RESULTS AND DISCUSSIONS

**Sample fabrication and theoretical description**

As a basis for the structuring procedure, a thin (111) YIG film with a thickness of $t$ = 44 nm is used, which is grown on top of a 500 μm thick (111) Gadolinium Gallium Garnet (GGG) substrate by Liquid Phase Epitaxy (LPE) (*25, 26*). A preliminary characterization by stripline Vector-Network-Analyzer ferromagnetic resonance (VNA-FMR) spectroscopy (*27, 28*) is performed to obtain the fundamental magnetic properties. The measurement, shown in Supplementary Fig. S1, yields a saturation magnetization of $M_s = (140.7 \pm 2.8) \frac{\text{kA}}{\text{m}}$ and a Gilbert damping parameter of $\alpha = (1.75 \pm 0.08) \times 10^{-4}$. These values are common for high-quality YIG thin films (*25, 26*). Thereafter, the nanostructuring process is carried out by utilizing a hard mask ion beam milling procedure (Supplementary Fig. S2) to fabricate various YIG waveguides with widths ranging from $w$ = 5 μm down to $w$ = 50 nm. Subsequently, a coplanar waveguide (CPW) antenna is fabricated on top of the waveguides which allows for the excitation of spin waves by applying a radiofrequency (RF) current. An in-detail description of the entire nanostructuring procedure is given in the Methods and Supplementary Sec. S2. Figure 1A shows scanning electron microscopy (SEM) micrographs of the smallest fabricated structure with $w$ = 50 nm.

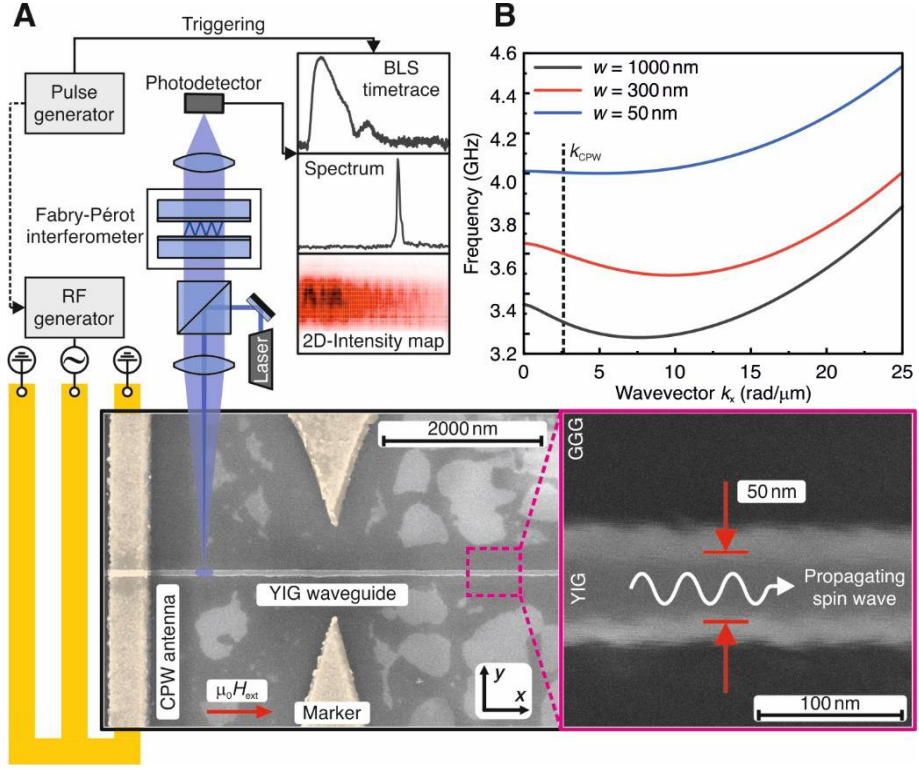

**Fig. 1. Schematic overview of the experimental configuration and SEM-micrographs of the smallest structure under investigation.** **(A)** The YIG waveguides are magnetized along the long axis (BV geometry) with an external magnetic field of $\mu_0 H_{ext}$ = 55 mT. A radio-frequency (RF) current is applied to a coplanar waveguide antenna (CPW) on top of the waveguide and propagating spin waves are excited. Frequency, spatial and time resolved scans are performed via micro-focused BLS. The width $w$ of the structure is defined using the visibly unharmed core, since it can be assumed that the magnetic properties of the edges are significantly alternated due to the exposure to the ion bombardment. **(B)** Theoretical BV spin-wave dispersion relations for YIG-waveguides with a respective width of $w$ = 1000 nm, 300 nm and 50 nm. (*24*). $k_{CPW}$ denotes the wavevector at which the CPW antenna possesses the largest excitation efficiency.

For the theoretical description of the system, the BV geometry spin-wave dispersion has to be considered (*24*):

$$\omega = \sqrt{\left(\omega_H + \left(\lambda^2 K^2 + F_{k_x}^{zz}\right)\omega_M\right) \times \left(\omega_H + \left(\lambda^2 K^2 + F_{k_x}^{yy}\right)\omega_M\right)}, \quad (1)$$

where $\omega_M = \gamma\mu_0 M_s$, $\omega_H = \gamma B = \gamma\mu_0 H_{ext}$ due to a negligible demagnetization along the $x$-direction and $\lambda = \sqrt{\frac{2A_{ex}}{\mu_0 M_s^2}}$ with the exchange constant $A_{ex}$ and the vacuum permeability $\mu_0$. Moreover, $F_{k_x}^{zz}$ and $F_{k_x}^{yy}$ denote the dynamic demagnetization tensor components out-of-plane ($z$-direction) and in-plane perpendicular to the waveguide ($y$-direction) and $K^2 = k_x^2 + k_y^2 + k_z^2$ is the total wavevector. In a confined system, the phenomenon of pinning takes place caused by the dipolar interaction which can be taken into account by introducing an effective width $w_{eff} > w$ of the structure (*29, 30*). Thus, considering only the fundamental width mode, $k_y$ can be written as $k_y = \frac{\pi}{w_{eff}}$. For a decreasing structure size, the contribution of the exchange interaction will increase and eventually dominate over the dipolar interaction. This will lead to an effective unpinning of the system described by the fact that $w_{eff}$ tends to infinity ($w_{eff} \to \infty$) beyond this so-called unpinning threshold, as it is shown in Supplementary Fig. S3. A comprehensive description of this matter was recently given in (*24*). It should be noted that, since the thickness of the investigated film is very small, the modes are always unpinned in $z$-direction, hence $k_z = \frac{p\pi}{t}$. Here, $p$ denotes the mode number of these so-called perpendicular standing spin-wave (PSSW) modes. To summarize:

$$K^2 = k_x^2 + \left(\frac{\pi}{w_{\text{eff}}}\right)^2 + \left(\frac{p\pi}{t}\right)^2. \tag{2}$$

The experimental investigation of the spin-wave dynamics is carried out using micro-focused BLS by focusing a laser beam on top of the waveguides and analyzing the inelastically scattered light, while a small external magnetic field $\mu_0 H_{\text{ext}}$ is applied in *x*-direction parallel to the long axis of the waveguide (BV geometry). Frequency, spatial, and time resolved measurements of the spin-wave intensity are performed (*31*) as it is schematically illustrated in Fig. 1A.

## Measurement of the thermal population

Since BLS is also sensitive to incoherent magnons, the thermal population of the magnon dispersion can be measured. Figure 2A shows such an exemplary measurement for a YIG waveguide of *w* = 1000 nm. Two peaks are detected which correspond to the fundamental waveguide mode and the first PSSW mode. Performing a field dependent measurement and using Eq. (1) to fit the frequency of the PSSW mode´s intensity maximum, with the approximation of $k_x = 0$, gives access to the exchange constant $A_{\text{ex}}$ as a fitting parameter. The results presented in Fig. 2B show no significant dependency on the structure width, which indicates an insignificant influence of the nanostructuring process. An average exchange constant of $A_{\text{ex}} = (4.22 \pm 0.21) \times 10^{-12} \frac{\text{J}}{\text{m}}$ can be extracted, which is within the typical range for YIG (*32, 33*). In addition, a decline of the PSSW mode intensity is observed which denies the extraction of the PSSW mode frequency below a width of *w* = 200 nm, which is further discussed in Supplementary Sec. S4.

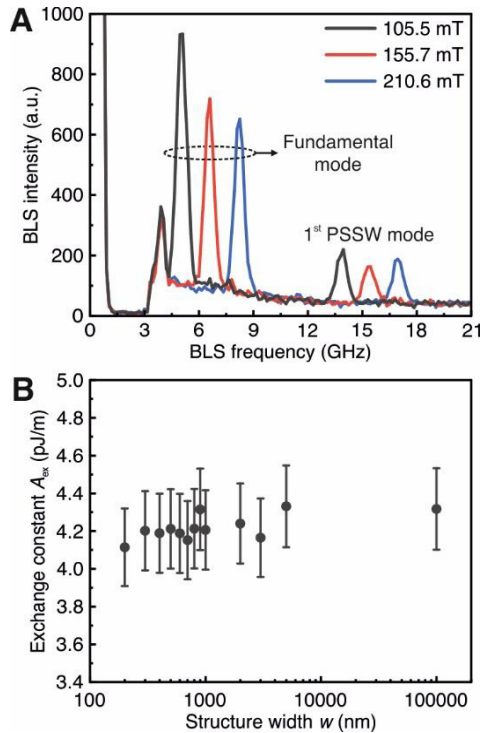

**Fig. 2. Measurement of the thermal spin-wave population and determined exchange constant. (A)** Exemplary thermal BLS spectra in the absence of any microwave excitation for a *w* = 1000 nm YIG waveguide. A field dependent measurement of the perpendicular standing spin-wave (PSSW) modes frequencies allows for the determination of the exchange constant according to Eq. (1) **(B)** Extracted exchange constant $A_{\text{ex}}$ in dependency of the structure width *w*. No significant dependency is found, which indicates an insignificant influence of the nanostructuring process.

Using the extracted exchange constant, the dispersion of the fundamental mode can be calculated based on Eq. (1), as it is exemplarily shown in Fig. 1B for the investigated waveguides with a respective width of *w* = 1000 nm, 300 nm and 50 nm. Furthermore, the expected group velocity $v_g$ can be calculated as the derivative of the dispersion and moreover, assuming an unchanged Gilbert damping constant, the expected decay length $\lambda_D$ can be derived. An overview for different investigated structure widths is given in Supplementary Fig. S5. In the following, these theoretical predictions are compared to the experimentally measured spin-wave spectra.

**Spin-wave spectra and decay length**

To acquire the respective spin-wave spectrum for each structure width, a RF current is applied to the CPW antenna to excite spin waves. Using micro-focused BLS, the spin-wave intensity is measured in the center of the waveguide close to the edge of the CPW. Sweeping the applied RF frequency $f_{ex}$ yields the spectrum, as it is exemplarily shown in Fig. 3A - C (solid black lines) for structure widths $w$ = 1000 nm, 300 nm and 50 nm. The theoretical spin-wave spectrum (dashed black lines in Fig. 3A - C) is calculated from the excitation efficiency of the CPW antenna (Supplementary Fig. S5) and the spin-wave dispersion. For larger structure sizes, a good qualitative agreement between theory and experiment is found. However, a frequency shift is observed, which increases for decreasing structure size (see Supplementary Fig. S7) to approximately 200 MHz for the smallest structures. A potential cause for this shift can be found in the increasing influence of laser heating for decreasing structure size. An influence of the structuring process cannot be excluded either. Additionally, in Fig. 3C an unexpectedly broad spectrum is observed. This is an indicator for a nonlinear behavior of the spin waves and the appearance of a nonlinear shift due to a high magnon density which would also lead to a downshift of the measured spectrum (*9, 34–36*).

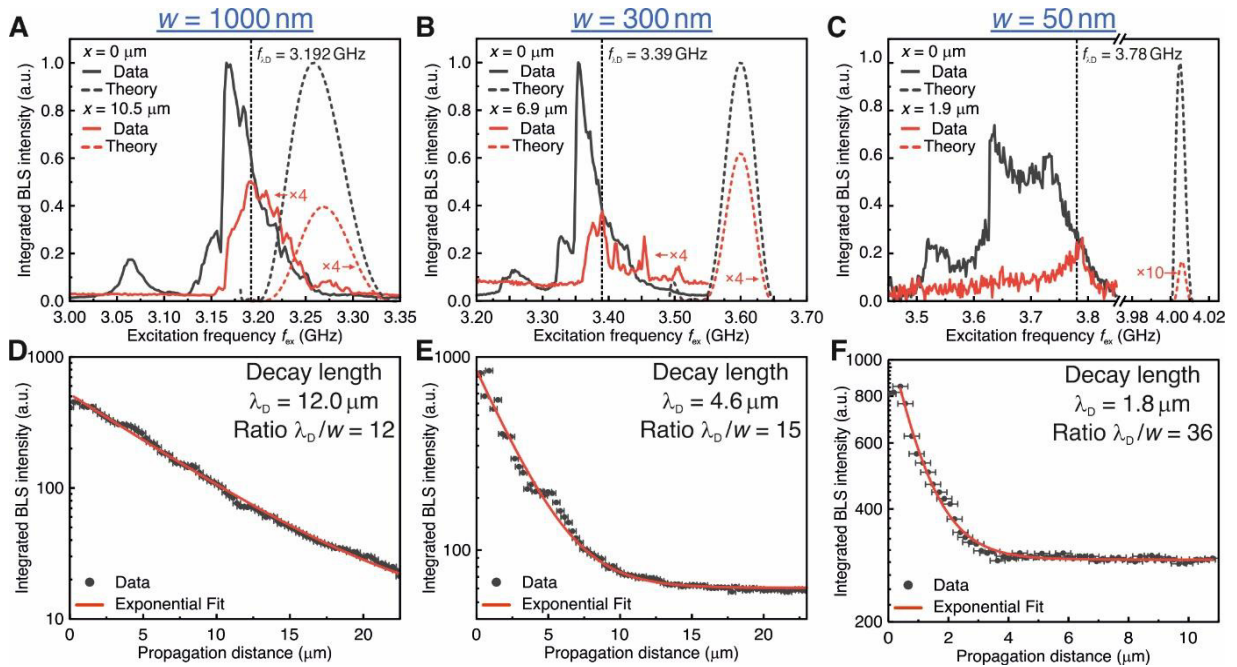

**Fig. 3. RF-excitation spectra and decay length measurements. (A) - (C)** Measurement (solid lines) in comparison to the theoretical calculation (dashed lines) of the excited spin-wave spectra close to the CPW (black) and in a certain distance to the CPW (red) for $w$ = 1000 nm, 300 nm and 50 nm and $\mu_0 H_{ext}$ = 55 mT. **(D) - (F)** Spin-wave intensity integrated across the width of the waveguide vs. the propagation distance along the waveguide. An exponential fit according to Eq. (3) yields the decay length $\lambda_D$. The excitation frequency is selected from (A) - (C) by choosing the spin waves which exhibits the highest intensity after a certain propagation distance. This frequency is marked by the vertical dashed black line in (A) - (C).

In the following, the decay length $\lambda_D$ is extracted from the experiment. This parameter is crucial considering any application of spin waves, since it defines the range of information transport and determines the energy consumption of devices.

Depending on the chosen excitation frequency and the excited mode, the decay length might vary due to a change in group velocity and lifetime. To select the frequency which possesses the largest decay length, a second spectrum (Fig. 3A - C solid red line) is acquired in a certain set distance to the edge of the CPW antenna. The intensity maximum of this spectrum is shifted to the frequency of the wave which has the best trade-off between velocity and propagation losses and thus, the largest decay length within the accessible excitation regime. This frequency $f_{\lambda D}$, marked by a vertical dashed black line in Fig. 3A - C, is used for further investigation. A spatial resolved 2D-scan across and along the waveguide is performed while spin waves with $f_{\lambda D}$ are continuously excited. Integrating the measured intensity across the width results in the decay graphs shown in Fig. 3D - F. To describe the observed decrease, an exponential decay according to

$$I(x) = I_1 \exp\left(\frac{2x}{\lambda_D}\right) + I_0 \tag{3}$$

can be fitted. Here, $I_1$ denotes the initial intensity, $x$ the position along the waveguide and $I_0$ the offset intensity due to the background of thermal spin waves. The results are displayed in Fig. 3D - F. For $w$ = 1000 nm the decay length is found to be $\lambda_D$ = (12.0 ± 1.2) µm. For smaller structure widths, a decrease to $\lambda_D$ = (4.6 ± 1.1) µm for $w$ = 300 nm down to $\lambda_D$ = (1.8 ± 0.4) µm for $w$ = 50 nm is observed. The appearance of nonlinear effects, as seen in Fig. 3C, might influence these measurements. This is due to the fact that they constitute an additional loss channel for the initial spin wave, hence the presented results can be understood as a lower limit estimation. In fact, the observed decay lengths already fulfill the fundamental requirement regarding the realization of nano-scaled magnonic logic circuits, which is the information conservation on the length scale of the respective logic gate. Moreover, the potential circuit complexity, which can be defined as the decay length over the feature size $R_{PCC} = \lambda_D/w$, increases from $R_{PCC}$ = 12 for $w$ = 1000 nm to $R_{PCC}$ = 36 for $w$ = 50 nm (see Fig. 3D,F and Supplementary Fig. S8). Hence, with decreasing structure size more advanced logic operations can be achieved without the need of intermediate amplification of the spin waves.

In the following, the measured decay lengths are compared to the theoretical expectation. The experimental results for all investigated structure widths are shown in Fig. 4A (black dots) and confirm the observed decline of the decay length. One can calculate the theoretical expectation (see Supplementary Fig. S5) under the assumption that the Gilbert damping parameter remained unchanged during the structuring process. For further comparison, the maximum of the theoretically expected decay length, within the accessible wavevector excitation regime, is extracted and plotted in Fig. 4A (red dots solid line). The theoretical values describe the results reasonably well, but are slightly smaller than the experimental data for larger structure widths above $w$ = 1000 nm. This discrepancy might be caused by the determination of the initial Gilbert damping parameter. Since the used method essentially probes a large sample area also inhomogeneities are included which might lead to an overestimation of the damping, whereas a local measurement in a nanostructure yields a damping closer to the real intrinsic value (19). Furthermore, the experimental data shows a quasi-saturation level below $w$ = 100 nm, whereas the theory decreases monotonously. A careful analysis has to be made to exclude that the observed saturation level is a measurement artifact, since the direct excitation by the CPW antennas far field can mimic an exponential decay. This influence is ruled out by theoretical estimations as discussed in Supplementary Sec. S5 and by extracting the group velocity from time resolved measurements, as discussed below.

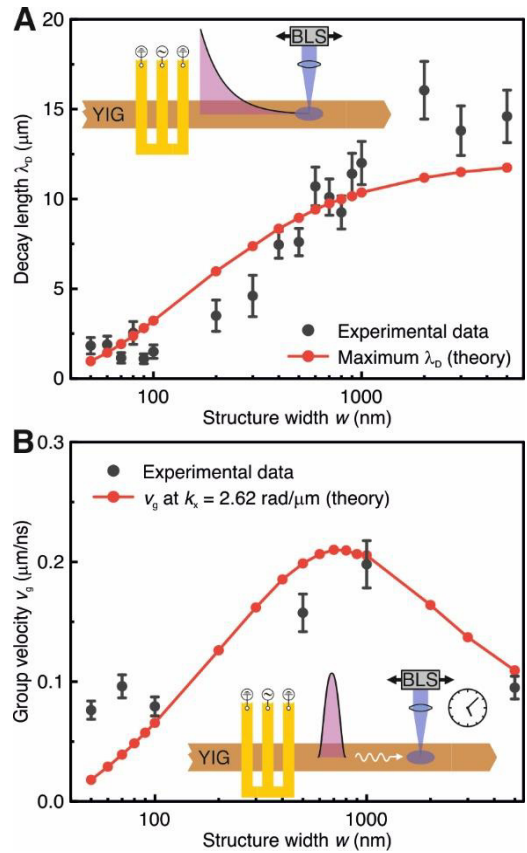

**Fig. 4. Measured decay length and group velocity in dependency of the structure width. (A)** The observed decay length decreases with decreasing structure width. Comparing to theoretical calculations, a reasonable agreement is found which indicates an intrinsic origin, found to be the group velocity, rather than a loss of quality during the structuring process. **(B)** This is verified by confirming the theoretical expectations for the group velocity by employing time-resolved micro-focused BLS.

## Spin-wave group velocity

Both, the theoretical calculation and the experimentally observed decay length, show a reasonable accordance, which indicates only a moderate influence of the structuring process since unchanged initial parameters of the YIG have been assumed for the calculations. Therefore, the cause of the decreasing decay length is of rather intrinsic nature and can be found in the group velocity $v_g$. The dipolar branch of the dispersion flattens successively with decreasing structure size, as it is visible in Fig. 1B and Supplementary Fig. 5. Thus, the group velocity, which is the derivative of the dispersion relation, decreases substantially and subsequently also the decay length (Supplementary Fig. S5). To experimentally validate this assumption, the group velocity is measured directly. By applying 50 ns long excitation pulses to the CPW antenna, spin-wave wave packets are excited. Utilizing time-resolved micro-focused BLS, the wave packet can be tracked in time and space, which gives access to the group velocity. A detailed description of the measurements is given in Supplementary Sec. S8. Figure 4B shows the resulting group velocity for several waveguide widths (black dots). A good qualitative agreement of experiment and theory is found confirming the theoretical assumptions. Only for structures with $w < 100$ nm the results lie above the expectation, which explains the quasi-saturation level of the decay length for those structures. Potentially, during the course of their propagation, the wavevector of the observed waves is shifted to the exchange regime which would result in an increased group velocity. This can be caused by a frequency downshift of the dispersion due to laser heating, whereas two-magnon-scattering provides the mechanism to transform the wavevector while the frequency of the spin wave is conserved. An exemplary calculation for $w = 50$ nm shows that only a small frequency shift of 28 MHz is necessary to cause a wavevector transformation from 2.62 rad/µm to 10.5 rad/µm (transformation of the wavelength from 2.4 µm to 600 nm) which corresponds to the observed group velocity. Nevertheless, these measurements are ultimately proving the propagation of spin waves in these nano-conduits, since any kind of direct excitation of the CPW antenna is immediately separated by the time resolution.

**CONCLUSION**

To conclude, we presented a study of the propagation of spin-wave packets in individual magnonic conduits with lateral dimensions down to 50 nm. Space and time resolved micro-focused Brillouin-Light-Scattering spectroscopy was used to extract the exchange constant and directly measure the spin-wave decay length and group velocity. The decrease of the decay length in dependency of the conduit width theoretically predicted by (24) was proven experimentally showing a decrease from $\lambda_D$ = (12.0 ± 1.2) µm for $w$ = 1000 nm down to $\lambda_D$ = (1.8 ± 0.4) µm for $w$ = 50 nm, which indicates only a moderate influence of the structuring process on the YIG parameters. The decrease is caused by a successive flattening of the dipolar branch of the spin-wave dispersion, due to the interplay of the in-plane and out-of-plane demagnetization tensor components leading from an elliptical to a circular spin precession when approaching an aspect ratio of 1. In spite of the drop in the free path, the potential circuit complexity increases by a factor of three from $R_{PCC}$ = 12 to $R_{PCC}$ = 36 rendering nano-waveguides more attractive for the construction of magnonic circuits. Surprisingly, the measured free path in a 50 nm wide waveguide is two times larger than the expected theoretical value of $\lambda_D$ = 0.97 µm which is, however, in agreement with the unexpectedly high spin-wave group velocity obtained from direct measurements. This indicates that fast exchange-dominated spin waves of short wavelengths down to 600 nm are responsible for the transfer of energy in the smallest waveguides rather than long-wavelength dipolar waves. The presented demonstration of spin-wave propagation on the nanoscale, comparable to the scale of modern CMOS, is a significant milestone on the way towards the development of novel magnonic circuits.

**MATERIALS AND METHODS**

**Sample fabrication**

A 44 nm thin LPE grown (111) YIG film on a 500 µm thick (111)-oriented GGG substrate was used for the fabrication of the nanostructures. Prior to the structuring process, the sample was cleaned by means of an ultrasonic bath in acetone and isopropanol. Afterwards an oxygen plasma treatment was used to remove any organic residuals. Ti/Au (20 nm/40 nm) alignment marker and structure labels were fabricated utilizing a Lift-Off technique with Electron Beam (E-Beam) Lithography and E-Beam Physical Vapor Deposition. YIG waveguides of 60 µm length with tapered ends and widths of $w$ = 5 µm, 3 µm, 2 µm 1 µm, 900 nm, 800 nm, 700 nm, 600 nm, 500 nm, 400 nm, 300 nm, 200 nm, 100 nm, 90 nm, 80 nm, 70 nm, 60 nm and 50 nm were fabricated using a hard mask ion beam milling procedure. This procedure was based on a Cr/Ti stack of 30 nm/15 nm thickness as the hard mask and successive $Ar^+$ ion milling under 20°, 70° and 20° incident angle with respect to the film normal. An additional in-detail description is given in Supplementary Sec. S2. In a final step ground-signal-ground (GSG) CPW antennas were structured on top of the waveguides made from Ti/Au (10 nm/80 nm). The respective width of the GSG lines is 400 nm-600 nm-400 nm with a G-S center-to-center distance of 1.1 µm.

**Micro-focused BLS spectroscopy**

The sample is placed in-between the poles of a large electromagnet to provide a variable and spatially homogeneous magnetic field. A continuous-wave single-frequency laser operating at 457 nm is focused through the substrate of the sample on the respective structure to be investigated using a compensating microscope objective (magnification 100x, numerical aperture NA = 0.85). The effective laser spot size is approximately 300 nm and the effective laser power on the sample is 2 mW. Analyzing the inelastically scattered light using a multi-pass Tandem-Fabry-Pérot interferometer and a single photon detector yields the frequency shift of the light and thus, due to momentum and energy conversation, the frequency of the magnons. The measured BLS intensity is proportional to the spin-wave intensity and in-plane spin-wave wavevectors up to 24 rad/μm can be detected. For the measurement of the thermal population, the absolute frequency resolution is 150 MHz. A piezoelectric driven nano-positioning stage allows for spatial resolved scans by moving the sample with respect to the objective and in addition, is used to realize an optical stabilization to compensate thermal drifts. Furthermore, using a pulse generator allows for the synchronization of the microwave excitation source and the detector output and thus, allows for time resolved measurements.

## SUPPLEMENTARY MATERIALS

Fig. S1. Results of the VNA-FMR spectroscopy.
Fig. S2. Schematically depicted nanostructuring process.
Fig. S3. Effective width.
Fig. S4. Degradation of the 1$^{st}$ PSSW mode intensity.
Fig. S5. Overview of the theoretical calculations.
Fig. S6. Far field excitation of the CPW antenna.
Fig. S7. Frequency mismatch of the spin-wave spectra.
Fig. S8. Potential circuit complexity.
Fig. S9. Exemplary group velocity measurement for $w$ = 1000 nm.

**Acknowledgements**

This research has been funded by the European Research Council project ERC Starting Grant 678309 MagnonCircuits, by the Deutsche Forschungsgemeinschaft through the project DU 1427/2-1, by the Collaborative Research Center SFB/TRR-173 "Spin+X" (Project B01) and by the Austrian Science Fund (FWF) through the project I 4696-N. B.H. acknowledges support by the Graduate School Material Science in Mainz (MAINZ). The authors thank Burkard Hillebrands for support and valuable discussions.

**Author contributions**

C.D. fabricated the used YIG film. B.H., B.L., A.M.F., D.B. and S.S. carried out the nanostructuring of the sample. B.L. and B.H. acquired the SEM micrographs. M.S and B.H. prepared the measurement setup and B.H. conducted all VNA-FMR and BLS measurements and carried out the evaluation. B.H. and Q.W. performed all theoretical calculations. B.H. drafted the manuscript with the help of A.V.C., T.B. and P.P. The study was supervised by A.V.C, T.B. and P.P. All authors contributed to the scientific discussion and commented on the manuscript.

# Supplementary Materials

# Propagation of spin-waves packets in individual nano-sized yttrium iron garnet magnonic conduits

Björn Heinz[1,2]*, Thomas Brächer[1], Michael Schneider[1], Qi Wang[1,3], Bert Lägel[4], Anna M. Friedel[1], David Breitbach[1], Steffen Steinert[1], Thomas Meyer[1,5], Martin Kewenig[1], Carsten Dubs[6], Philipp Pirro[1] and Andrii V. Chumak[1,3]

[1]Fachbereich Physik and Landesforschungszentrum OPTIMAS, Technische Universität Kaiserslautern, D-67663 Kaiserslautern, Germany.
[2]Graduate School Materials Science in Mainz, Staudingerweg 9, D-55128 Mainz, Germany.
[3]Faculty of Physics, University of Vienna, Boltzmanngasse 5, A-1090 Wien, Austria.
[4]Nano Structuring Center, Technische Universität Kaiserslautern, D-67663 Kaiserslautern, Germany.
[5]THATec Innovation GmbH, Augustaanlage 23, D-68165 Mannheim, Germany.
[6]INNOVENT e.V., Technologieentwicklung Jena, D-07745 Jena, Germany.

*Corresponding author
E-Mail: bheinz@rhrk.uni-kl.de

**S1 – Preliminary characterization**

A preliminary characterization of the YIG film in use is carried out by means of stripline Vector-Network-Analyzer ferromagnetic resonance (VNA-FMR) spectroscopy in an in-plane magnetized configuration. A detailed description of this method can be found in (*27, 28*), here only the applied fitting equations and the results are stated:

$$f_{\mathrm{FMR}} = \bar{\gamma}\sqrt{\mu_0 H_{\mathrm{ext}}(\mu_0 H_{\mathrm{ext}} + \mu_0 M_{\mathrm{s}})}\,,\qquad\text{(S1)}$$

$$\mu_0 \Delta H = \mu_0 \Delta H_0 + \frac{2\alpha f_{\mathrm{FMR}}}{\bar{\gamma}}\,.\qquad\text{(S2)}$$

The influence of any internal in-plane magnetocrystalline anisotropy is neglected as it is expected to be rather weak (*25*). The measurement, depicted in Supplementary Fig. S1, yields a saturation magnetization of $M_{\mathrm{s}} = (140.7 \pm 2.8)\,\frac{\mathrm{kA}}{\mathrm{m}}$, an effective gyromagnetic ratio of $\bar{\gamma} = \frac{\gamma}{2\pi} = (28.17 \pm 0.56)\,\frac{\mathrm{GHz}}{\mathrm{T}}$, an inhomogeneous linewidth broadening of $\mu_0 \Delta H_0 = (0.18 \pm 0.01)\,\mathrm{mT}$ and a Gilbert damping parameter of $\alpha = (1.75 \pm 0.08) \times 10^{-4}$.

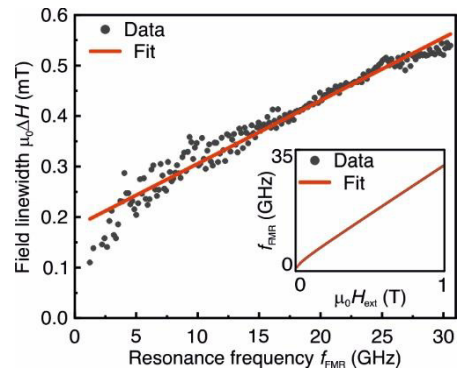

**Fig. S1. Results of the VNA-FMR spectroscopy.** A linear fit according to Eq. S1 of the field linewidth vs. the resonance frequency yields a Gilbert damping parameter of $\alpha = 1.75 \times 10^{-4}$. **Inset:** Resonance frequency vs. external magnetic field. Using Eq. S2 the saturation magnetization of $M_s = 140.7\,\frac{kA}{m}$ is extracted.

## S2 – Nanostructuring procedure

Here, an in-detail description of the nanostructuring process is given which is based upon a hard-mask ion beam milling procedure and is depicted schematically in Supplementary Fig. S2. Prior to the actual structuring of the YIG film, the sample is cleaned in an ultrasonic bath using acetone and isopropanol and is exposed to an oxygen plasma to remove organic residuals. Then, a double layer of photoresist, consisting of a bottom layer of PMMA AR-P 669.04 (dark blue layer) and a top layer of PMMA AR-P 679.02 (light blue layer), is applied using spin coating, where both layers exhibit a different molecule chain length (Panel A). Using Electron-Beam (E-Beam) Lithography the desired structures are written into the resist, which alternates the chemical structure of the exposed areas. Due to the different molecule chain length the top layer is left with a smaller alternated area than the bottom layer (Panel B). In a development step these areas are chemically removed and a resist mask with a so-called undercut is generated, which is necessary to guarantee the success of the Lift-Off process later on (Panel C). Using E-Beam Physical Vapor Deposition a stack of Cr (30 nm)/Ti (15 nm) is deposited on top of the whole sample (Panel D). In a Lift-Off step the resist is removed by acetone and consequently, the metal which is on top of the resist. Only the previously exposed areas remain covered by the metal and a so-called hard mask is generated (Panel E). In a successive multi-step ion beam milling process, the sample is bombarded with Ar$^+$ ions under incident angles of 20°, 70° and 20° with respect to the film normal (Panel F). The unprotected YIG, as well as the Ti layer of the mask, are removed by the ion bombardment, and only the desired YIG structures and some residual Cr on top of these remain. Subsequently, the Cr is removed by a wet etch process using an acid that YIG is inert to (Panel G), resulting in the final structure (Panel H).

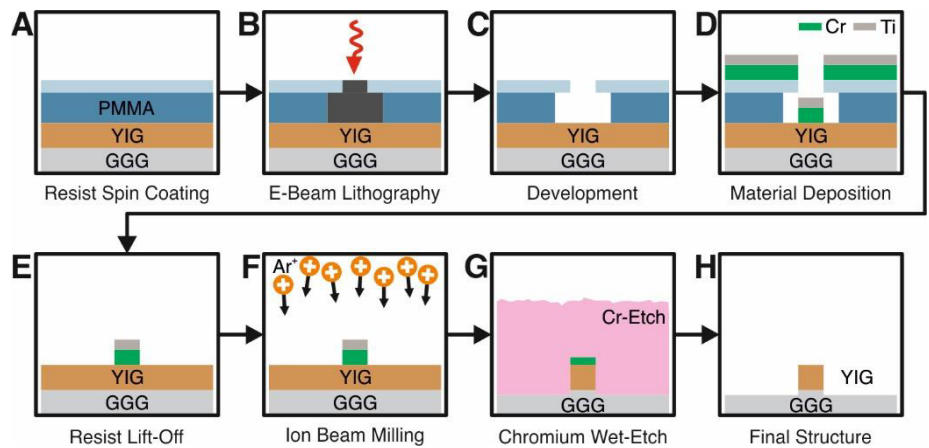

**Fig. S2. Schematically depicted nanostructuring process. (a-c)** A resist mask with an undercut is generated by E-Beam Lithography. **(d-e)** Deposition of Cr(30nm)/Ti(15nm) by means of E-Beam Physical Vapor Deposition and a successive Lift-Off result in a hard mask. **(f-h)** After Ar$^+$ ion milling and removal of the residual hard mask the final structure is achieved.

Although this hard mask procedure is a quite time consuming and complex method, it is superior compared to common microstructuring processes directly utilizing a resist mask as a protective layer for a milling step. The reason for this lies within the difficulty to fabricate high aspect ratio (height to width ratio) masks, which is necessarily the case if the intended structure size decreases to the nanometer regime. By utilizing Titanium, which exhibits a much higher etch resistance than any polymer resist mask, much thinner masks can be used. Furthermore, using Chromium as an additional seed layer allows for an easy removal of the residual mask afterwards. It should be noted that this procedure is especially suitable for YIG but might not be utilized for metallic ferromagnets, since it is tricky to find an acid with the proper selectivity regarding Chromium and the magnets constituents.

## S3 – Effective width and unpinning threshold

Typically, the phenomenon of pinning, caused by the dipolar interaction, is taken into account by introducing an effective width $w_{eff} > w$ (*29, 30*). This width corresponds to the imaginary width at which a full pinning of the dynamic precession occurs. However, the exchange interaction counteracts this tendency and eventually dominates if the width of the respective element is decreased, forcing the system into a fully unpinned state. This is shown in Supplementary Fig. S3 using the inverse effective width $w/w_{eff}$, which can be understood as a pinning parameter (*24*). If the system is fully pinned this ratio will be equal to 1, whereas the fully unpinned state equals 0. For the investigated waveguides the unpinning threshold width is approximately $w_{crit}$ = 220 nm, marked by the vertical dashed black line in Supplementary Fig. S3.

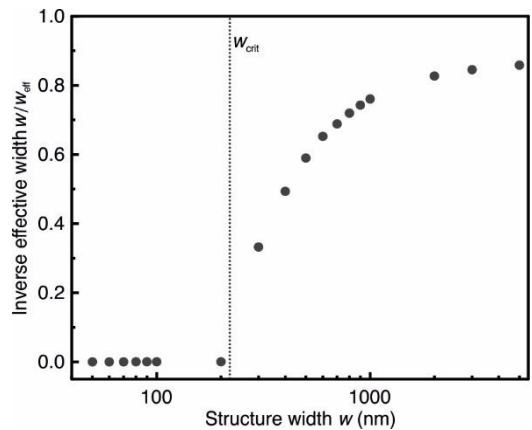

**Fig. S3. Effective width.** Calculated inverse effective width for all investigated structures vs. the structure´s width to show the pinning condition. The system is always fully unpinned below the unpinning threshold width of $w_{crit}$ = 220 nm, marked by the vertical dashed black line.

## S4 – Thermal BLS measurements

As it can be seen in Supplementary Fig. S4 the observed 1st PSSW mode intensity decreases with decreasing structure size. The potential origin can be linked to multiple effects. Especially the BLS sensitivity is changed with decreasing structure size, since the effectively probed volume decreases drastically for structures smaller than 300 nm (size of the laser spot). In addition, the effective scattering cross section has a complex dependency on the mode profile and thus on the width of the structure. Furthermore, the slight trapezoidal shape of the structures might lead to the observed distortion of the mode, since the thickness is no longer well defined, but alternated. Moreover, a potential structural damage due to the nanostructuring process cannot be ruled out.

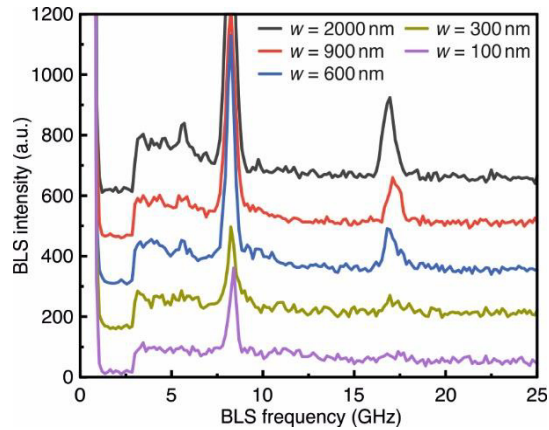

**Fig. S4. Degradation of the 1st PSSW mode intensity.** Thermal BLS spectra for different structure widths w and an external field of $\mu_0 H_{ext}$ = 211.2 mT. A clear drop of the PSSW intensity is observed for decreasing structure size whereas the fundamental mode remains unchanged. The spectra are vertically shifted with respect to each other for better visibility.

### S5 – Theoretical calculations

In this chapter, an overview of the theoretical calculations is given. The dispersion, group velocity and decay length for different waveguide widths *w* shown in Supplementary Fig. S5A - C are calculated under the assumption of unchanged initial parameters of the YIG and using the measured exchange constant (Fig. 2). Moreover, the amplitude excitation efficiency of the CPW antenna is depicted in Supplementary Fig. S5D, which is derived as the Fourier transformation of the total magnetic field distribution (Inset Supplementary Fig. S6). For a detailed derivation of the magnetic field, see Reference (*37*). It should be noted that only the out-of-plane (oop) field has to be taken into account, since the in-plane field cannot directly excite spin waves in the used measurement configuration. Furthermore, the shown excitation efficiency is only proportional to the real excitation efficiency since the effective mode profile and the ellipticity of the precession further influence the coupling (*38*). Therefore, a relative comparison between the signal strength of different structures is not possible.

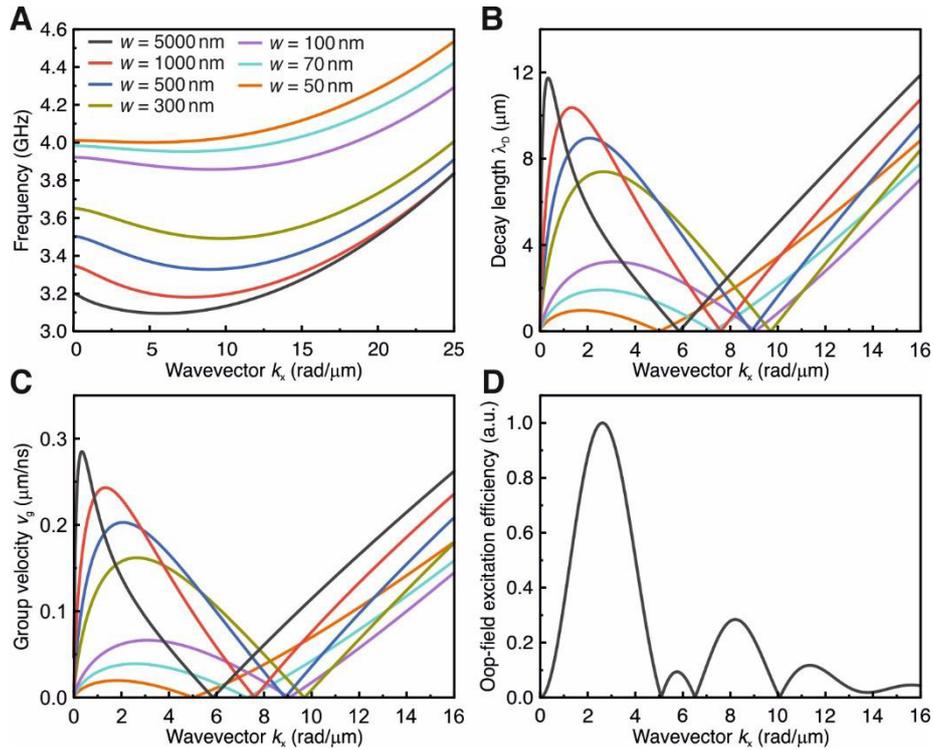

**Fig. S5. Overview of the theoretical calculations. (A)** Dispersion, **(B)** Decay length under the assumption of an unchanged Gilbert damping parameter and **(C)** Group velocity for an external field of $\mu_0 H_{ext}$ = 55 mT. **(D)** Calculated out-of-plane (oop) field amplitude excitation efficiency of the CPW antenna in use.

Since the direct excitation by the CPW antennas far field is independent of the structure size and can mimic an exponential decay, a careful analysis has to be made to exclude that the measurements of the decay length are distorted by this effect. To estimate this influence, the decay of the calculated oop-field is fitted with an exponential decay as shown in Supplementary Fig. S6, which yields a small decay length of only $\lambda_D$ = 300 nm. Thus, the far field's influence on the measurements is expected to be minor.

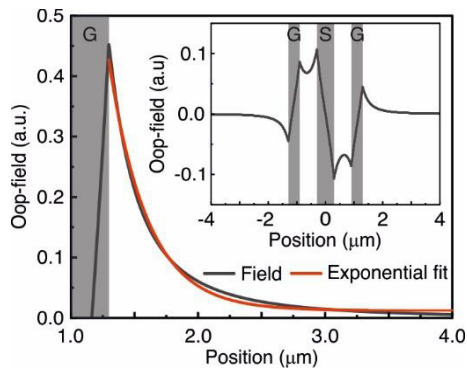

**Fig. S6. Far field excitation of the CPW antenna.** An exponential fit of the oop-field yields a small decay length of 300 nm due to the field confinement of the CPW configuration, hence the influence on the measurements is expected to be minor. **Inset**: Total field distribution of the oop-field. The shaded areas mark the CPW antenna.

### S6 – Frequency mismatch

Supplementary Fig. S7 shows the frequency of the measured and calculated spin-wave spectra. The frequency mismatch increases with decreasing structure size to approximately 200 MHz for w = 50 nm.

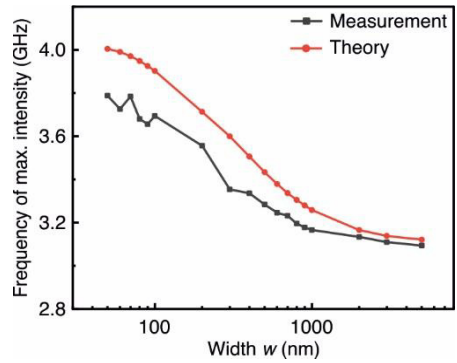

**Fig. S7. Frequency mismatch of the spin-wave spectra.** Frequency of the maximum intensity of measured and calculated spin-wave spectra for all investigated waveguide widths $w$.

### S7 – Potential circuit complexity

The ratio $R_{PCC} = \lambda_D/w$ can be understood as a normalization of the decay length with the feature size of the respective structure and is used to define the so-called potential circuit complexity. This is based on the following consideration: Assuming, based on the feature size, a grid with $w \times w$ cell size, a larger potential circuit complexity allows for a longer wiring within a single decay length and thus allows for more complex logic operations without the need of additional amplification. In Supplementary Fig. S8 the potential circuit complexity for all investigated structure widths is shown.

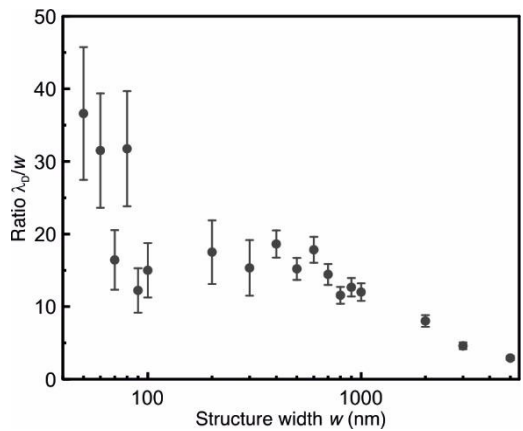

**Fig. S8. Potential circuit complexity.** An increasing potential circuit complexity is observed for decreasing structure size.

### S8 – Measurement of the group velocity

The group velocity is measured by exciting spin-wave wave packets in a pulsed manner and tracking a distinct spin-wave mode along the waveguide by micro-focused time resolved BLS spectroscopy. To do so, a pulsed excitation spectrum is acquired close to the CPW antenna by applying 50 ns long microwave pulses with varying carrier frequency and the frequency of the intensity's maximum is determined and selected for further investigation. Then, the time resolved spin-wave intensity is measured in the center of the waveguide along the propagation direction, as depicted in Supplementary Fig. S9. The time corresponding to the respective maximum intensity is defined as the peak arrival time. Performing a linear fit of the measurement position vs. the peak arrival time, as shown in the Inset in Supplementary Fig. S9 yields the group velocity of the respective spin-wave mode. For comparison to the theoretical calculations, the group velocity corresponding to the wavevector of the CPW antennas maximum excitation efficiency is assumed to be the wavevector associated with the maximum intensity of the frequency spectrum.

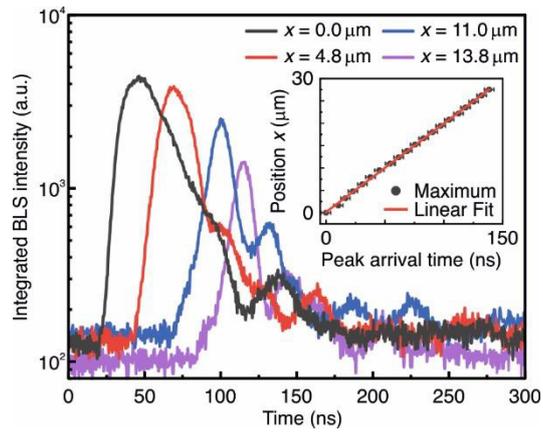

**Fig. S9. Exemplary group velocity measurement for *w* = 1000 nm.** Spin-wave wave packets are excited by 50 ns excitation pulses and their time traces are measured at different positions along the waveguide. The peak arrival time is defined by the maximum intensity which is measured. **Inset:** A linear fit of the measurement position vs. the peak arrival time yields the group velocity of the traced mode.